\documentclass{PoS}
\usepackage{amsmath,amssymb}

\pdfoutput=1
 
\newlength{\ppno}
\newlength{\pplace}
\pplace=\linewidth

\newcommand{\preprint}[1]{
 \settowidth{\ppno}{\rm\normalsize #1}
 \addtolength{\pplace}{-\ppno}
 \begin{picture}(0,0)
    \put(\pplace\unitlength,132){\rm\normalsize #1}
 \end{picture}
}

\renewcommand{\phi}{\varphi}
\newcommand{\smallfrac}[2]{\mbox{\small ${\displaystyle \frac{#1}{#2}}$}}
\newcommand{\x}{\tilde x}

\title{Modified Lattice Landau Gauge\\
        \preprint{ADP-07-16/T656}}

\ShortTitle{Modified Lattice Landau Gauge}

\author{\speaker{Lorenz von Smekal}\hspace{-2pt}, \hspace{-1pt}
        Dhagash Mehta, Andr\'e Sternbeck\thanks{Supported by the
        Australian Research Council}
        \ and Anthony G. Williams\\ %  
        Centre for the Subatomic Structure of Matter, School of Chemistry \& Physics\\ The University of Adelaide, Adelaide, SA 5005, Australia\\
        E-mail: \email{lorenz.smekal@adelaide.edu.au}}

\abstract{We propose a modified lattice Landau gauge based on
  stereographically projecting the link variables on the circle
  $S^1\to \mathbb{R}$ for compact $U(1)$ or the 3-sphere $S^3\to
  \mathbb{R}^3$ for $SU(2)$ before imposing the Landau gauge
  condition. This can reduce the number of Gribov copies exponentially
  and solves the Gribov problem in compact $U(1)$ where it is a
  lattice artifact. Applied to the maximal Abelian subgroup this might
  be just enough to avoid the perfect cancellation amongst the Gribov copies
  in a lattice BRST formulation for $SU(N)$, and thus to avoid the Neuberger
  0/0 problem. The con\-tinuum limit of the Landau gauge remains unchanged.}

\FullConference{The XXV International Symposium on Lattice Field Theory\\
		 July 30 - August 4 2007\\
		 Regensburg, Germany}

\begin{document}

\section{Introduction}

%Within the language of 
The machinery for
dealing with redundant degrees of freedom due to gauge invariance 
in local quantum field theory is based on  
%Becchi-Rouet-Stora-Tyutin (BRST) 
BRST symmetry. %which %is a global symmetry and 
%can be considered the quantum version of local gauge invariance. 
In QED it reduces to the long-known Gupta-Bleuler
construction. In non-Abelian gauge
theories, all is well in perturbation theory also. Beyond that,
however, there is a problem with such constructions that has 
not been fully and comprehensively addressed as yet. It relates to the
famous Gribov ambiguity, 
%\cite{Gribov1978},
the existence of Gribov copies that satisfy the Lorenz
condition (or any other local gauge fixing condition) but are related
by gauge transformations, and are thus physically equivalent. 
%As a result of this ambiguity, the usual definitions of a BRST charge fail
%to be globally valid.   
Lattice gauge theory provides a rigorous non-perturbative framework
whose strength and beauty derives from the fact that gauge invariance
is manifest and fixing a gauge is not required. However, in order to
arrive at a non-perturbative definition of non-Abelian gauge theories
in the continuum, from a lattice formulation, we need to be able to
perform the continuum limit in a formally watertight way.  
When attempting to fix a gauge via BRST formulations on the lattice,
the Gribov ambiguity manifests itself in the notorious Neuberger
problem which asserts that the expectation value of any gauge
invariant (and thus physical) observable in a lattice BRST 
formulation will always be of the indefinite form 0/0
\cite{Neuberger1987}. 

%Significant progress in extended lattice BRST
%formulations without this problem has been made recently [28--31].

One possible way forward % to avoid this problem 
%might be 
is provided by the 
ghost/anti-ghost symmetric Curci-Ferrari gauges. While the Neuberger
0/0 problem can be extended to include these non-linear gauges 
with their extended double-BRST symmetry despite their quartic ghost
self-interactions, they do allow the introduction of a mass term for
ghosts. Such a Curci-Ferrari mass breaks the nilpotency 
of the BRST/anti-BRST charges which is known to result in a loss
of unitarity and which perhaps therefore meant that this relatively old model
received little attention for many years. On the other hand, this mass
also serves to regulate the Neuberger zeroes in a lattice formulation. 
Expectation values of observables can then be meaningfully defined 
in the limit $m \to 0$ via l'Hospital's rule 
\cite{Kalloniatis2005,Ghiotti2006,vonSmekal2007}.

Here we propose an alternative more directly relating to the methods
of topological quantum field theory on the gauge group
\cite{Baulieu1998,Schaden1998}.   
The problem there is that the partition function of the topological model
to be used as the gauge-fixing device computes a topological invariant
such as the Euler characteristic of the gauge group which vanishes. In the
Landau gauge, for example, this partition function reduces to the
sign-weighted sum over all Gribov copies as the critical points of a
Morse potential on the gauge orbit which corresponds to computing the
Euler characteristic of the gauge group via the Poincar\'e-Hopf
theorem. This is the limit of vanishing gauge parameter, $\xi \to
0$. Neuberger's 0/0 argument is based on the opposite limit, $\xi \to
\infty$. Even though there is no gauge-fixing in this limit in the
first place, the topological model does not depend on $\xi$ and
actually computes the same invariant. This can be seen most clearly in
the ghost/anti-ghost symmetric Curci-Ferrari model for $SU(N)$, where
in the massless case the gauge-fixing partition function for $\xi \to
\infty$ factorises into a product of one Gauss-Bonnet integral
expression for the vanishing Euler characteristic $\chi(SU(N))$ per each
site of the lattice \cite{vonSmekal2007}. 

The 0/0 problem due to the vanishing Euler characteristic of $SU(N)$  
is avoided when fixing the gauge only up to the maximal Abelian
subgroup $U(1)^{N-1}$ because the Euler characteristic of the coset
manifold is non-zero. In Ref.~\cite{Schaden1998} the corresponding
lattice BRST has been explicitly constructed for $SU(2)$, where the coset
manifold is the 2-sphere and $\chi(SU(2)/U(1)) = \chi(S^2) = 2$. 

Is this an indication that the Neuberger problem might be solved when
that of compact $U(1)$ is? The cancellation of Gribov copies is
certainly present there already because $\chi(S^1) = 0$. Here we
propose a perhaps surprisingly simple solution to this problem by
stereographically projecting the circle $S^1 \to \mathbb{R}$ which can be
achieved by a simple modification of the minimising potential. This
then leads to a positive (semi)definite Faddeev-Popov operator, and
there is thus no cancellation of Gribov copies in compact $U(1)$ with
this so modified lattice Landau gauge.

The same procedure can be readily extended and applied to $SU(2)$ 
with the 3-sphere as group manifold and projecting $S^3 \to
\mathbb{R}^3$.
Even though we don't have an explicit proof for $SU(2)$ (except in one
dimension) that this solves the Neuberger problem, the remaining
manifold relevant in the topological argument then becomes
$\mathbb{R}P^2$ whose Euler characteristic is unity. For a lattice
BRST construction based on this topological invariant we certainly do
not expect that there are no contributions from any Gribov copies outside
the first Gribov horizon (with both signs of the Faddeev-Popov
determinant), as we know there are such copies in the 
continuum limit. The topological argument indicates, however,
that once the cancellation in a one-parameter $U(1)$-subgroup
direction along the gauge orbits is avoided, there will be no perfect
cancellation anymore between remaining copies of different signs, as
$\chi(\mathbb{R}P^2)=1$.  
%This is what we can achieve. The $SU(2)$ Faddeev-Popov operator of our
%modified Lattice Landau gauge is (semi)positive along $U(1)$
%directions. 
%Whether or not that is sufficient to avoid the Neuberger
%problem, as we are confident that it is, it is important to note that
%because our proposed modifications do not affect the continuum limit,
%one implementation of the lattice Landau gauge is as well jsutified as
%the other as far as the continuum gauge theory is concerned. Any
%observable differences between the two are lattice artifacts. Luckily,
%as we will see by comparison, these differences appear to be surprisingly
%small.
Whether that is really sufficient to avoid the Neuberger
problem or not, it is important to note that our %proposed
modifications do not affect the continuum limit. One implementation of
lattice Landau gauge is as well justified as the other as far as
the continuum gauge theory is concerned. Any observable differences
between the two are lattice artifacts. Luckily, as we will see by
comparison, these differences appear to be surprisingly small.
 
\section{Modified lattice Landau gauge for compact $U(1)$}

To be specific, for compact $U(1)$ with link angles $\phi_{i,\mu} \in
(-\pi,\pi] $ and gauge transformations  $\phi_{i,\mu} \to
  \phi_{i,\mu}^\theta = \phi_{i,\mu} + \theta_{i+\mu} - \theta_i \mod
  2\pi$ by gauge angles $\theta_i \in (-\pi,\pi]$ per site $i$ of the
    lattice, consider in parallel the $2\pi$-periodic potentials on
    gauge orbits $\phi $ defined by 
\begin{equation}
V_\varphi [\theta] = \sum_{i,\,\mu} ( 1-\cos\phi_{i,\mu}^\theta
) \; ,
\;\; \mbox{and} \quad \widetilde V_\varphi [\theta] = -2
\sum_{i,\,\mu} \ln\big((1+\cos\phi_{i,\mu}^\theta)/2\big) \; .
\vspace{-4pt}
\end{equation}
The first one is that of standard lattice Landau gauge corresponding
to the height function on the circle per link as a proper Morse
potential for the $U(1)$ gauge group on the lattice. Its critical
points are the Gribov copies of standard lattice Landau gauge. In the
modified potential $\widetilde V_\phi $ we have introduced a
singularity to suppress the South pole. Variation w.r.t.~the gauge
angles $\theta_i$ gives, 
\begin{equation}
f_i(\theta)  = - \sum_{\mu} ( \sin\phi_{i,\mu}^\theta -
\sin\phi_{i-\mu,\mu}^\theta ) \; , 
\;\; \mbox{and} \;\;\; \tilde f_i(\theta)  = -2 \sum_{\mu} \big(
\tan(\phi_{i,\mu}^\theta/2) -  \tan(\phi_{i-\mu,\mu}^\theta/2) \big)
\; . \label{gauge_cond}
\vspace{-7pt}
\end{equation}
The solutions to the Landau gauge conditions $f_i = 0 $ or $\tilde f_i =
0$, for $i=1,\dots (\# \mathrm{sites}) $, define the set of Gribov copies
in either case. To see the cancellation of Gribov copies in the
standard case explicitly, consider the admittedly trivial example of
compact $U(1)$ in 1 dimension as a simple warm-up exercise. Of
course, there is no dynamics in 1 dimension as there is essentially
only a single gauge orbit. Nevertheless, the number of solutions to
the standard Landau gauge condition grows exponentially with the
problem size. To see this, consider a finite chain with $n$ sites. To
remove global gauge transformations, impose anti-periodic boundary
conditions on the link and gauge angles ({\it i.e.}, $C$-periodic
b.c.'s for the corresponding group elements). With periodic b.c.'s the
Faddeev-Popov matrix will, of course, always have one exact zero
eigenvalue due to the unfixed global gauge invariance. This one
remaining zero is otherwise inessential for the present argument. With
a.p.b.c.'s the solutions to the conditions $f_i=0$, $i=1,\dots n$, are
then found to be simply given by sequences of link angles where every link
$\phi_i^\theta $ can only either be $0$ or $\pi$, no matter what the
initial set of link angles $\phi_i$ is. There are $2^n$ such
configurations which are Gribov copies of the trivial configuration
$\phi =0 $ ({\it i.e.}, $\phi_i=0$, $i=1,\, \dots n$). This is because
the number of links equals the number of sites, in 1 dimension. With
a.p.b.c.'s, every link configuration is gauge-equivalent to $\phi=0$ 
(and gauge-equivalent up to a constant with p.b.c.'s).
In the corresponding gauge-fixing
partition function of standard lattice BRST, 
\begin{equation} 
 Z_\mathrm{GF} \,=\, \int_{V^n} \smallfrac{d^n\theta}{(2\pi)^n} 
 \int_{\mathbb{R}^n} d^nb \;\,
 \mathrm{det}\, M_\mathrm{FP}  \; \exp\big\{i(b,f)-\smallfrac{\xi}{2}
 (b,b) \big\} \, \stackrel{\xi = 0}{=} \, \int_{V^n} d^n\theta
 \;\,  \mathrm{det}\, M_\mathrm{FP} \;\delta^n(f(\theta))   \;,  
\end{equation}
where $V^n$  %=\{\theta\in \mathbb{R}^n:\, \}$ 
is the integration range of the gauge angles $ -\pi \le \theta_i \le
\pi $ and $M_\mathrm{FP} $ is the Faddeev-Popov matrix,
%$i=1,\,\dots n$, 
%and ``$(,)$'' stands for the scalar product in $\mathbb{R}^n$, is 
we can therefore perform a variable transform $\theta \mapsto s = \phi
+ M \theta$, where $(M\theta)_i = \theta_{i+1} -\theta_i$. 
With a.p.b.c.'s,  $|\mathrm{det}\, M | = 2$ reflecting the fact that the
map is 2 to 1 when applying modulo $2\pi  $ on $s$ (changing every
gauge angle by $\pi$ leaves the link configuration unchanged mod
$2\pi$). With periodic b.c.'s the matrix $M$ has one eigenvalue zero
as we can't change the average value of the link angles by a gauge
transformation. With this variable transform, we see that in
the anti-periodic case the $2^n$ solutions $s^{(k)}$ (with  components
$s^{(k)}_i = 0$ or $\pi $) are obtained from $\phi= 0$ with $s^{(k)} = 
M\theta^{(k)}$. Inverting this relation, we find two classes of
solutions for $\theta \in V^n$: (a) sequences with all $\theta_i $ in
$\{0,\,\pi\}$ and (b) sequences with all $\theta_i $ in
$\{\pi/2,-\pi/2\}$. There are thus together $2^{n+1} $ solutions to
$f=0$ for the gauge angles $\theta$ in $V^n$ (the double counting of
link configurations with $2\pi$-periodicity corresponds to identifying
pairs of $\theta$, related by a swapping of all components,
in each of the 2 types separately). Type (a) is easily seen 
to always lead to a link configuration $s^{(k)}$ with an even number
of $\pi$'s, while type (b) always leads to on odd number of links in
the South pole.  We have, 
\begin{equation} 
 Z_\mathrm{GF} \,= \, %\int_{I(V^n)} d^ns \, |\mathrm{det}\, M|^{-1} 
 %\,  \mathrm{det}\, M_\mathrm{FP} \;\delta^n(f(\theta(s))) \,= \,
 2 \int_{V^n} d^ns \; |\mathrm{det}\, M|^{-1} 
  \,  \mathrm{det}\, M_\mathrm{FP} \;\delta^n\big(f(\theta(s))\big) \,= \,
  2 \, \sum_{k=1}^{2^n} \mathrm{sign}\big(\mathrm{det}\,
    M_\mathrm{FP}(s^{(k)})\big) \; .  \vspace{-2pt}
\end{equation}
The Faddeev-Popov matrix is obtained from
\begin{equation}
  (M_\mathrm{FP})_{ij} \, = \, \smallfrac{\partial
  f_i(\theta)}{\partial\theta_j}\,  =  \, -\cos s_i \, \delta_{i+1,j}\,
  + \,( \cos s_i + \cos s_{i-1}) \, \delta_{i,j}\, - \,\cos s_{i-1} \,
  \delta_{i-1,j} \; , 
\end{equation}
where we have used the variables $s_i = \phi_i + \theta_{i+1} -
\theta_i$ as above. It can be written in the form
\begin{equation}
M_\mathrm{FP} \,= \, M^T D(\cos s_1, \, \cos s_2, \,\dots\,  \cos s_n)\, M
\; , \vspace{-2pt}
\end{equation}
where $D(c_1, \, c_2, \, \dots\, c_n)$ is a diagonal  $n\times n$ matrix
with entry $c_i$ at position $i$ along the diagonal. The real matrix
$M$ for a.p.b.c.'s is the same non-singular matrix as above. From
Sylverster's law of inertia, the numbers of positive, negative and
zero eigenvalues of the Hermitian Faddeev-Popov matrix
$M_\mathrm{FP}$  are the same as those of the diagonal matrix $D$. For
the Gribov copies $s^{(k)}$  of our $U(1)$ chain with all $s^{(k)}_i =
0$ or $\pi $ there are no zero eigenvalues. Moreover, we see that we
have exactly one negative eigenvalue for every $\pi$ link in the chain.  
The number of $\pi$'s is even and the Faddeev-Popov determinant thus
positive for all type (a) solutions, and negative for those of type
(b) with an odd number of $\pi$ links.  The number of copies 
of type (a) and (b) are the same, each with the $2^n$ gauge configurations
$\theta $ in $V^n$ yielding  $2^{n-1}$ inequivalent link configurations mod
$2\pi$. Therefore,
\begin{equation} 
 Z_\mathrm{GF} \,= \,   2 \, \sum_{k=1}^{2^n} \mathrm{sign}\big(\mathrm{det}\,
    M_\mathrm{FP}(s^{(k)})\big)  \, = 2 \, \big( 2^{n-1} \, -\,
    2^{n-1} \big) \, = \, 0 \; . 
\end{equation}
This is the Neuberger zero. Observe the $Z_2$ structure: we can pick
any one solution of type (b) to define a gauge transformation which when
applied to any copy of type (a) or (b) leads to a copy of the other
type, and returns the original configuration when applied twice.
For odd $n$ for example, the checkerboard gauge transformation
$\theta = \big(\smallfrac{\pi}{2}, -\smallfrac{\pi}{2}, \,\dots\, 
\smallfrac{\pi}{2}\big)$ changes every link by $\pi$ and therefore
swaps the overall sign of $M_\mathrm{FP}$. Because $n$ is odd, this
then also swaps the sign of the Faddeev-Popov determinant (the same
works with periodic b.c.'s and even $n$ for $\theta = (0,\pi,0,\dots
\, \pi)$; because there is one zero eigenvalue in the periodic case,
the number of non-zero ones for even $n$ is odd and their product then 
changes sign also). It is very easy to see that this twofold symmetry
amongst Gribov copies of opposite sign in fact exists in the standard
Landau gauge for compact $U(1)$ in any dimension, and with it the
Neuberger 0/0 problem.   
 
For our anti-periodic 1-dimensional model, the modified Landau gauge
condition in Eq.~(\ref{gauge_cond}), stereographically projecting 
the links $s_i \to \tan(s_i/2)$,  now analogously leads to a
Faddeev-Popov matrix of the form%\\[-18pt] 
\begin{equation}
\widetilde M_\mathrm{FP} \,= \, %\smallfrac{1}{2} \; 
  M^T D\big(\sec^2 \smallfrac{s_1}{2}, \, \sec^2  \smallfrac{s_2}{2},
  \,\dots\,  \sec^2  \smallfrac{s_n}{2} \big)\, M
\; . %\vspace{2pt}
\end{equation}
Again, with a.p.b.c.'s, the real matrix $M$ is non-singular, while with
periodic b.c.'s it has one zero eigenvalue. The diagonal elements of
$D$ are all positive (in fact $\ge 1$) and so is
$M_\mathrm{FP}$. Moreover, the Faddeev-Popov matrix in higher
dimensions consists of a sum of terms of the same form, one for every
dimension. As a sum of positive (semi-)definite matrices it remains
such in any dimension. 

The twofold symmetry among the solution that changes the sign of the
Faddeev-Popov determinant no-longer exists. In 1 dimension, of the
originally $2^{n+1}$  solutions only 2 remain. There are 
no type (b) solutions to $\tilde f = 0$, and  of type (a) only $\theta
= (0, \, 0, \, \dots \, 0 )$ and $(\pi, \, \pi, \, \dots \, \pi )$
survive which both lead to the same trivial link configuration mod
$2\pi$. There are thus no more Gribov copies. 
This changes in higher dimensions, where there are copies which, {\it
  e.g.}, in 2 dimensions differ from each other by a number of vortex pairs. 
Nevertheless, the Faddeev-Popov matrix is (semi-)positive for {\em any}
choice  of $\theta $, not only for the solutions to $\tilde f =
0$. This means that the  potential $\widetilde V_\phi[\theta] $ is
convex to the above. It does have singularities, but there is always a
unique minimum lying in between the infinite walls. The gauge-fixing
partition function counts their number, {\it i.e.}, the 
Gribov copies in this modified Landau gauge, which should not depend
on the gauge orbit. This may sound strange at first, but it 
is quite possible in compact $U(1)$ where the copies (all minima here)
can be classified entirely in terms of their vortex content
\cite{Akino2002}. It will 
not be the case for $SU(N)$ generalisations.\footnote{Neither does it 
  have to be because of the partial cancellations that then still
  occur, {\it e.g.}, in computing $\chi(\mathbb{R}P^2$).}   

Note that the singularities in  $\widetilde V_\phi[\theta] $ are
crucial in order to escape Neuberger's argument. Consider the
partition function of our modified model with an additional parameter
$t$ as in \cite{Neuberger1987},  
\begin{align} 
\widetilde Z_\mathrm{GF}(t) 
 \, &= \, \int_{V^n} \smallfrac{d^n\theta}{(2\pi)^n}  
  \int_{\mathbb{R}^n} d^nb \;\,
 \mathrm{det}\, \big(t \widetilde M_\mathrm{FP}\big) 
 \; \exp\Big\{it (b,\tilde f)-\smallfrac{\xi}{2} (b,b) \Big\}
  \\
 &= \, t^n \int_{V^n} \smallfrac{d^n\theta}{(2\pi\xi)^{n/2}\hspace*{-.4cm}}  
  \hskip .5cm
 \mathrm{det}\, \widetilde M_\mathrm{FP}  
 \; \exp\Big\{-\smallfrac{t^2}{2\xi} (\tilde f,\tilde f) \Big\} 
\, = \, t^n \sum_\mathrm{copies} 
\int_{\mathbb{R}^n} \smallfrac{d^n\tilde
 f}{(2\pi\xi)^{n/2}\hspace*{-.4cm}}     \hskip .5cm
\exp\Big\{-\smallfrac{t^2}{2\xi} (\tilde f,\tilde f) \Big\} \; .  \nonumber
%\, = \, N_\mathrm{copies} \; .
\end{align}
Because $\mathrm{det}\,
\widetilde M_\mathrm{FP} $ is always positive, it compensates the
Jacobian of our variable transform $\theta \to \tilde f$.   
If $\tilde f$ was bounded, we could let $t\to 0$ and, from the
prefactor $t^n$, observe the Neuberger zero. We can't do that anymore
without encountering an infinity $\propto t^{-n}$ from the Gaussian
integration over $\tilde f$ because of the stereographic projection of
the link angles in $\tilde f$  which decompactifies the integration
range. The result is of course still $t$-independent and simply given
by the number of Gribov copies,
\begin{equation} 
\vspace{-2pt}
      \widetilde Z_\mathrm{GF}(t) \, = \, (\#\mathrm{copies}) \;\;
       [\, = \, 2\, , \; \mbox{in $1$ dim.}\,]\,  ,
      \quad \mbox{independent of $t$}\; . 
\vspace{-2pt}
\end{equation}
This thus solves the Neuberger problem of compact $U(1)$ in any dimension.

\section{Stereographically projected lattice Landau gauge for $SU(2)$}

\begin{figure}[t]
  \centering
  \begin{minipage}[t]{0.48\textwidth}
    \includegraphics[width=\linewidth,height=6.2cm]{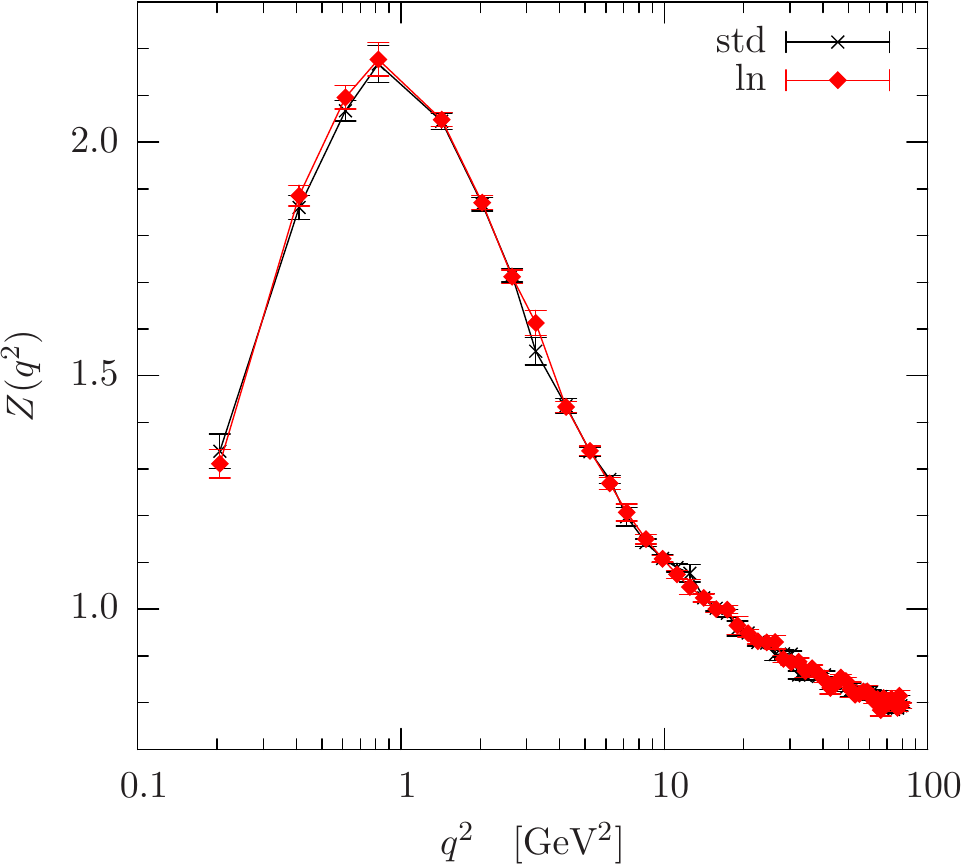}
  \end{minipage} 
  \hfill
  \begin{minipage}[t]{0.48\textwidth}
    \includegraphics[width=\linewidth,height=6.2cm]{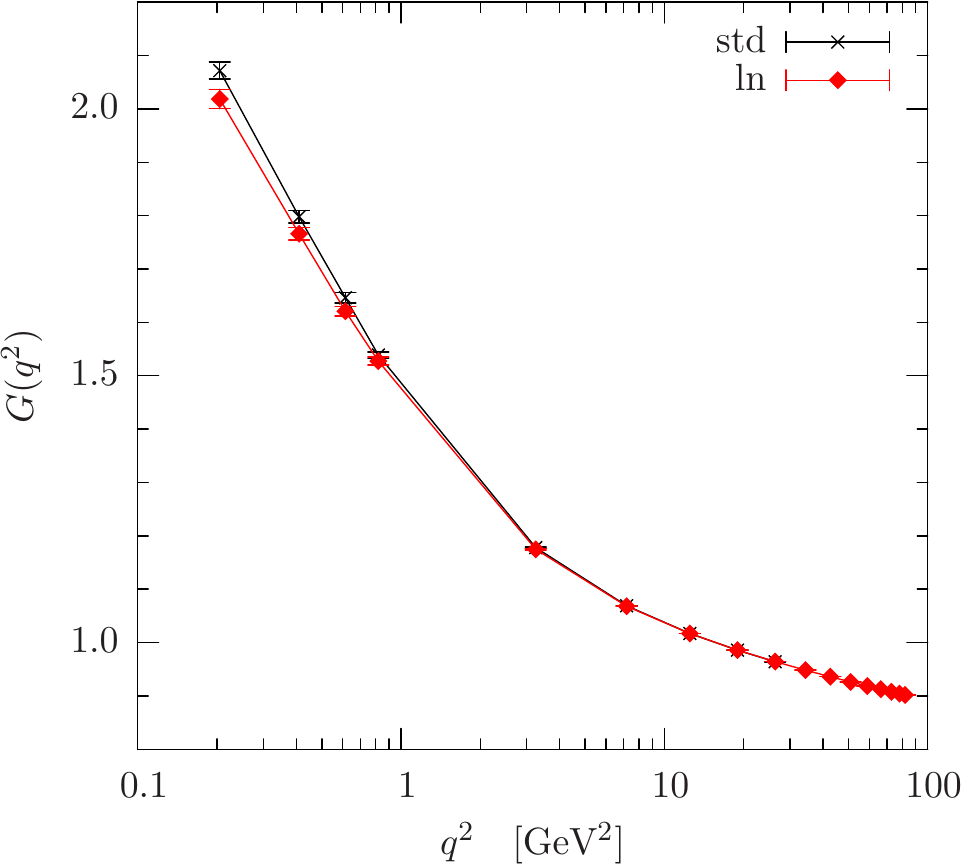}
   \end{minipage}
    \caption{The gluon (left) and ghost (right) dressing functions 
      for pure SU(2) gauge theory at $\beta = 2.5$ on a
      $32^4$ lattice (renormalised at $\mu=4$~GeV) with standard (std)
      and modified (ln) lattice Landau gauge-fixing.}
    \vspace*{-.2cm}
    \label{gl_gh_dress}  
\end{figure}

In order to analogously suppress the South pole, in the minimising
potential for $SU(2)$, we essentially have to replace 
$\frac{1}{2} \, \mbox{tr}\;  U^g_{i,\mu}  \, \to \, \ln\big(1+
\frac{1}{2} \, \mbox{tr}\;  U^g_{i,\mu} \big) $ where $U^g_{i,\mu} =
g_i^\dagger U_{i,\mu} g_{i+\mu}$ is the gauge transformed $\mu$-link
at $i$, with $U, g\in SU(2)$. As above we thus compare the
gauge-fixing potentials,
\begin{equation}
V_U [g] = \smallfrac{2}{\rho}  \sum_{i,\,\mu} \Big( 1-  \smallfrac{1}{2}
 \; \mbox{tr}\;  U^g_{i,\mu}   \Big) \; ,
\;\; \mbox{and} \quad \widetilde V_U [g] = -  \smallfrac{4}{\rho} 
\sum_{i,\,\mu} \ln\Big(\smallfrac{1}{2}+\smallfrac{1}{4}
 \; \mbox{tr}\;  U^g_{i,\mu} \Big) \; .
 \label{min_pot_SU2}
\vspace{-4pt}
\end{equation}
With $U^g = x^0 + i \sigma^a x^a $, $(x^0,\vec x) \in S^3$, and
stereographically projected  $\tilde x^a = x^a/(1+x^0)$ as the
corresponding variables in $\mathbb{R}^3$, the respective Landau gauge
conditions become, 
\begin{equation}
f_i^a  \, = - 2\, \sum_{\mu} \, ( x_{i,\mu}^a - x_{i-\mu,\mu}^a ) \; , 
\;\; \mbox{and} \quad 
\tilde f_i^a \,  = - 4 \,\sum_{\mu}\, ( \tilde x_{i,\mu}^a -
\tilde x_{i-\mu,\mu}^a ) \; . \label{gauge_cond_SU2}
\vspace{-8pt}
\end{equation}
The modified Faddeev-Popov operator can then be obtained in the form,
with $\x^0=x^0/(1+x^0)$,
\begin{align}
  (\widetilde M_\mathrm{FP})_{ij}^{ab} = 2\sum_\mu \Big\{
  - &\big( \x_{i,\mu}^0
   \delta^{ab} + \epsilon^{abc} \x_{i,\mu}^c + \x^a_{i,\mu}\x^b_{i,\mu}\big)
   \delta_{i+\mu,j}   
  +  \big(( \x_{i,\mu}^0 + \x_{i-\mu,\mu}^0) \delta^{ab} +
  \x^a_{i,\mu} \x^b_{i,\mu}  \nonumber \\ 
  &+ \x_{i-\mu,\mu}^a \x_{i-\mu,\mu}^b \big) \delta_{ij} 
    - \big( \x_{i-\mu,\mu}^0 \delta^{ab} - \epsilon^{abc} \x_{i-\mu,\mu}^c +
   \x^a_{i-\mu,\mu}\x^b_{i-\mu,\mu} \big) \delta_{i-\mu,j}   \Big\} \;
  . \label{mod_M_FP}
 %\nonumber
\vspace{-8pt}
\end{align}
The new feature, due to the logarithm in (\ref{min_pot_SU2}), are the
$\x^a \x^b$-terms quadratic in the projected variables $\x$. Apart from those,
$\widetilde M_{FP}$  is of the same form as the standard 
Faddeev-Popov operator in $SU(2)$ with the $x$'s replaced by  $2\x$.
The stereographically projected variables have the property that 
$\x^0 + \x\cdot\!\x  =\frac{1}{2}\, \sec^2\frac{\theta}{4}$, where
$\frac{1}{2} \, \mbox{tr}\;  U^g   = x_0 =\cos  \frac{\theta}{2}$
defines the azimuthal angle of the link in $SU(2)$.  
This can be used to show that  $\widetilde M_\mathrm{FP}$ in (\ref{mod_M_FP})
is still positive semi-definite when  all links $\x_i^a $ are aligned
in the same $SU(2)$ direction. While this is true for all 
solutions to $\tilde f^a_i =0 $ in (\ref{gauge_cond_SU2}) in 1
dimension, it is not generally the case, and it does therefore
\emph{not} follow that in higher dimensions   $\widetilde M_\mathrm{FP}$ 
is generally a sum of positive terms, as it did in compact $U(1)$.  
  The principal advantage of the stereographic projection is that we do
not expect a perfect cancellation amongst Gribov copies of opposite
sign as mentioned in the introduction. This would then allow to define
a BRST formulation on the lattice.  

The main point here of course is that we can eventually include all
copies, not only minima, in a BRST-like average when using
stereographic projection, which would produce a zero  
result in the standard case. This would then be analogous to the
gauge-fixing procedure used in most functional continuum methods. 
This is not what we have done here. To assess the effect of the
modification, as a first step, we calculated the dressing
functions in the gluon and ghost propagators obtained 
from minimising each of the two potentials in (\ref{min_pot_SU2}). 
The results are compared in Fig.~\ref{gl_gh_dress}.

For the modified Landau gauge we used a specially adapted 
version of the Fourier-accelerated gauge fixing of \cite{Davies1987}.
The standard results and other details are those of \cite{SU2vsSU3}
where we used overrelaxation. We calculated the gluon and ghost
propagators on the same 50 starting configurations. After
minimisation, however, the gauge-fixed configurations of one potential
%(with maximal violations of the corresponding gauge condition below
%$10^{-14}$) 
still show considerable violations of the other potential's
condition. They are typically not nearly acceptable as
solutions to the other gauge condition. 
%not even approximately. 
Nevertheless, there are hardly any differences 
between the propagators obtained in each case. Even the differences
observed in the ghost propagator at low momenta are so small that they
might well be due to the different algorithms. The smallness of the
effect might in fact seem surprising given that we invert the
Hessian of different functions on different minima along the
gauge orbits, but it is necessary for the equivalence of the
two in the continuum limit. %, this is rather a non-trivial test.

%effect might in fact seem rather surprising given that we invert the
%Hessian of different functions on different minima along the
%gauge orbits.  Even though it is necessary for the equivalence of the
%two in the continuum limit, this is rather a non-trivial test.

\vspace{.4cm}

%\section{Conclusions}
\leftline{\large\bf Conclusions}

%\vspace{-8pt} 
\vspace{.1cm}

Stereographic projection to decompactify link variables before
gauge fixing solves the Neu\-ber\-ger 0/0 problem in compact $U(1)$. We have
good reason to expect it to prevent the cancellation of Gribov copies
similarly in $SU(2)$. Generalisation to $SU(N)$ may not be entirely trivial but
should also be possible. It is important to stress that the
modifications do not affect the gauge group. The proposal is to set
up a topological model that computes an invariant different from the
vanishing Euler characteristic of the gauge group.  This can be achieved
by mapping the South pole to infinity to suppresses parts of a
particular gauge orbit which should be of no harm but which is what
fixing a gauge is all about.  The continuum limit of the modified
lattice Landau gauge remains unchanged.
%, and the numerical results confirm
%this. 

%\vfill
\vskip .25cm

\noindent
%\textbf{Acknowledgement:} 
\textbf{ACK:} 
Discussions with M.~Schaden, 
Ph.~de~Forcrand  and D.~Zwanziger were greatly appreciated. 

%This research was supported
%by the Australian Research Council.


\vspace{-5pt} 


{\small
\begin{thebibliography}{99}
\addtolength{\itemsep}{-6pt}

\bibitem{Neuberger1987}
  H.~Neuberger, Phys.\ Lett.\ B {\bf 175} (1986) 69; {\it ibid.}
  {\bf 183} (1987) 337.

\bibitem{Kalloniatis2005}
  A.~C.~Kalloniatis, L.~von Smekal and A.~G.~Williams,
  %``Curci-Ferrari mass and the Neuberger problem,''
  Phys.\ Lett.\ B {\bf 609} (2005) 424 
  [\href{http://www.arXiv.org/abs/hep-lat/0501016}{hep-lat/0501016}].
  %\texttt{hep-lat/0501016}.
  %%CITATION = HEP-LAT 0501016;%%

\bibitem{Ghiotti2006}
  M.~Ghiotti, L.~von Smekal and A.~G.~Williams,
  %``Extended double lattice BRST, Curci-Ferrari mass and the Neuberger
  %problem,'' for {\em Quark Confinement and
  %the Hadron Spectrum VII}, Ponta Delgada, Azores, Portugal, September 2006, 
  AIP Conf.\ Proc.\  {\bf 892} (2007) 180 %[hep-th/0611058]. 
  [\href{http://www.arXiv.org/abs/hep-th/0611058}{hep-th/0611058}].
  %\texttt{hep-th/0611058}.
  %%CITATION = APCPC,892,180;%%

\bibitem{vonSmekal2007}
  L.~von Smekal, M.~Ghiotti and A.~G.~Williams, ``Decontracted double
  BRST on the lattice,'' in prep. 

\bibitem{Baulieu1998}
  L.~Baulieu and M.~Schaden, {Int.\ J.\ Mod.\ Phys.}\  {\bf A13}
  (1998) 985 %[hep-th/9601039].
  [\href{http://www.arXiv.org/abs/hep-th/9601039}{hep-th/9601039}].
  %\texttt{hep-th/9601039}. 
  %%CITATION = HEP-TH 9601039;%%

\bibitem{Schaden1998}
  M.~Schaden, {Phys. Rev.} D \textbf{59} (1998) 014508 %[hep-lat/9805020].
  [\href{http://www.arXiv.org/abs/hep-lat/9805020}{hep-lat/9805020}].
  %\texttt{hep-lat/9805020}. 
  %%CITATION = HEP-LAT 9805020;%%

\bibitem{Akino2002}
  See, {\it e.g.}, N.\ Akino and J.\ M.\ Kosterlitz, Phys.\ Rev.\ B\
  {\bf 66} (2002) 054536, and the references therein. 

\bibitem{Davies1987}
  C.~T.~H.~Davies {\it et al.},
  %``FOURIER ACCELERATION IN LATTICE GAUGE THEORIES. 1. LANDAU GAUGE FIXING,''
  Phys.\ Rev.\  D {\bf 37} (1988) 1581.
  %%CITATION = PHRVA,D37,1581;%%

\bibitem{SU2vsSU3}
  A.~Sternbeck, L.~von Smekal, D.~B.~Leinweber, and A.~G.~Williams,
  \pos{PoS(LATTICE 2007)340}.

\end{thebibliography}
}
\end{document}